\documentclass[apj]{emulateapj}
\slugcomment{{\sc Accepted for publication in the ApJ}}

\newcommand{\xsrc}{\mbox{SGR~1806--20~}}
\newcommand{\xsrcnos}{\mbox{SGR~1806--20}}
\newcommand{\sgrc}{\mbox{SGR~1900$+$14~}}

\begin{document} 
\vspace{0.8 in} 

\title{Extended Tails from \xsrc Bursts} 

\author{
 Ersin {G\"o\u{g}\"u\c{s}}\altaffilmark{1}, 
 Peter~M.~Woods\altaffilmark{2},
 Chryssa Kouveliotou\altaffilmark{3}
 Mark~H.~Finger\altaffilmark{4},
 Valentin Pal'shin\altaffilmark{5,6}
 Yuki Kaneko\altaffilmark{1},
 Sergey Golenetskii\altaffilmark{5},
 Dmitry Frederiks \altaffilmark{5},
 Carol Airhart\altaffilmark{7}
}
 
\altaffiltext{1}{Sabanc\i~University, Faculty of Engineering and Natural Sciences, Orhanl\i$-$Tuzla, \.Istanbul 34956 Turkey}
\altaffiltext{2}{Corvid Technologies, 689 Discovery Drive, Huntsville, AL 35806, USA}
\altaffiltext{3}{Space Science Office, VP-62, NASA/Marshall Space Flight Center, Huntsville, AL 35812, USA}
\altaffiltext{4}{Universities Space Research Association, 6767 Old Madison Pike, Huntsville, AL 35805, USA}
\altaffiltext{5}{Ioffe Physical-Technical Institute of the Russian Academy of Sciences, 26 Polytekhnicheskaya, St Petersburg 194021, Russia}
\altaffiltext{6}{St Petersburg State Polytechnical University, Politechnicheskaya 29, 195251 St Petersburg, Russia}
\altaffiltext{7}{Dynetics Inc., 1000 Explorer Boulevard, Huntsville, AL, 35806, USA}

\begin{abstract} 

In 2004, \xsrc underwent a period of intense and long-lasting burst activity that included the giant flare of 27 December 2004 -- the most intense extra-solar transient event ever detected at Earth. During this active episode, we routinely monitored the source with Rossi X-ray Timing Explorer and occasionally with Chandra. During the course of these observations, we identified two relatively bright bursts observed with Konus$-$Wind in hard X-rays that were followed by extended X-ray tails or afterglows lasting hundreds to thousands of seconds. Here, we present detailed spectral and temporal analysis of these events observed about 6 and 1.5 months prior to the 27 December 2004 Giant Flare. We find that both X-ray tails are consistent with a cooling blackbody of constant radius. These spectral results are qualitatively similar to those of the burst afterglows recorded from SGR 1900+14 and recently from SGR 1550$-$5418. However, the latter two sources exhibit significant increase in their pulsed X-ray intensity following the burst, while we did not detect any significant changes in the RMS pulsed amplitude during the \xsrc events. Moreover, we find that the fraction of energy partitioned to the burst (prompt energy release) and the tail (afterglow) differs by an order of magnitude between SGR 1900+14 and \xsrcnos. We suggest that such differences can be attributed to differences in the crustal heating mechanism of these neutron stars combined with the geometry of the emitting areas.

\end{abstract} 

\keywords{pulsars: individual (\xsrc) $-$ X-rays: bursts}

\section{Introduction} 
 
Soft Gamma Repeaters (SGRs) are a small group of isolated neutron stars that exhibit extraordinary properties. They are persistent X-ray sources with luminosities ranging between 10$^{33}$ and 10$^{35}$ erg s$^{-1}$, spin periods $2-9$ s and large spin down rates indicating extremely high surface dipolar magnetic field strengths of B$\sim$10$^{14}$$-$10$^{15}$ G.
They are distinguished from the overall neutron star population by their repeated emission of short and intense hard X-ray / soft gamma ray bursts with a typical duration of $\sim0.1$ s. The burst repetition timescale is of the order of a few seconds with burst peak luminosities, $L_{\rm peak}$, ranging between $10^{36}$-$10^{41}$ erg s$^{-1}$ \citep{gogus99, gogus00}. The burst fluence distribution is continuous and can be well fitted with a power law of index of around -1.6 \citep{gogus99,gogus00}. Consequently, although the short and less energetic events constitute the bulk of the SGR burst population, several intermediate size events (durations up to tens of seconds, $L_{\rm peak} \sim 10^{42}$ erg s$^{-1}$) have also been recorded in the last twelve years \citep{mazets99,kouv01,feroci03,mereg09}. Finally, SGRs emit giant flares, which are exceptionally rare bright events ($10^{44} < L_{\rm peak} <$ few$\times 10^{47}$ erg s$^{-1}$) that last several hundreds of seconds. Only three such events have been recorded to date; their durations, temporal structure and energetics are very similar (for a detailed review, see Woods \& Thompson 2006 or Mereghetti 2008).

An emerging class of SGR bursts referred to as intermediate bursts are always
associated with afterglows (tails) when observed with sensitive, low-background
X-ray detectors.  Lenters et al. (2003) compared three such bursts from \sgrc along with the giant flare of 27 August 1998, and estimated that in all four cases the
energy emitted during the afterglow (2$-$10 keV) was $\sim 2$\% of the total energy of the burst (25$-$100 keV). Their detailed spectral analysis of the tail
portion of two of these events revealed a decaying thermal component,
possibly associating the tails with the cooling of a burst-induced heating
process on the surface of the neutron star \citep{ibra01,lent03}. Unfortunately, there was no sensitive soft X-ray coverage of either the 1979 March 5 Giant Flare from SGR $0526-66$ or the June 1998 intermediate bursts from SGR $1627-41$. Recently, intermediate bursts from SGR 1550$-$5418 (aka 1E 1547.0$-$5408) were observed with multiple instruments during its 2009 burst active episode \citep{mereg09, kaneko10, eneto10}.

One of the most prolific sources, \xsrc, entered an active phase in March 2004 during which it emitted more than a thousand bursts over the following several months including the energetic Giant Flare of 2004 December 27 which had a peak luminosity of a few$\times$$10^{47}$ erg s$^{-1}$  \citep{hurl05,fred07}. We searched our {\it Rossi X-ray Timing  Explorer} (RXTE) observations of the source between January 2004 $-$ May 2005  and identified only two intermediate type events out of a total of over 1500 bursts. Each was followed by an extended X-ray tail or afterglow lasting several hundred seconds.  We were very fortunate to obtain simultaneous coverage of one of these bursts with {\it Chandra X-ray Observatory} (CXO). We  present here a detailed temporal and spectral analysis of the main burst and the afterglows of these two intermediate events (Sections 2 and 3), discuss their similarities and differences with other magnetar afterglows, and the implications thereof in Section 4. 

\section{Konus-Wind, RXTE and CXO Observations of the Two Intermediate Bursts}

\subsection{2004 June 22 event}

\begin{figure}[!h]
\vspace{-0.7in}
\centerline{
\includegraphics[scale=.5]{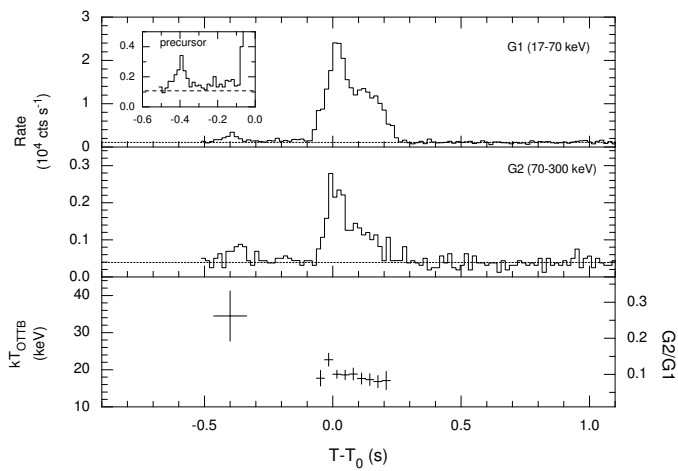}}
\vspace{-0.2in}
\caption{Konus-Wind light curve of the 2004 June 22 intermediate burst in the 17$-$70 keV band (top panel), and 70$-$300 keV band (middle panel). A close-up view of the precursor at T$_0$-0.460~s is shown in the inset of the top panel. Spectral evolution of the burst emission in terms of evaluated temperature of an OTTB model from the hardness ratio, G2/G1 (bottom panel). \label{fig:kw_ib1}}
\end{figure}

({\it i}) Konus-Wind: A relatively strong burst triggered the Konus-Wind instrument on 2004 June 22 at T$_0$=19:30:09.587 UT (Golenetskii et al. 2004a). The burst started with a rather weak precursor at T$_0$$-$0.460~s, followed by the main pulse at T$_0$$-$0.080~s. The T$_{90}$\footnotemark \footnotetext{The duration over
which 90\% of the burst emission has been recorded.} duration of the burst determined in the 17$-$300~keV range was 0.422$\pm$0.058~s (the total burst duration was about 0.726~s). 
We present the Konus-Wind light curve in the 17$-$70 keV (G1) and 70$-$300 keV (G2) energy bands in the top two panels of Figure~\ref{fig:kw_ib1}. We also present in Figure \ref{fig:kw_ib1} (bottom panel) the spectral evolution during the course of the burst in terms of the evaluated temperature of optically thin thermal bremsstrahlung (OTTB) obtained using the background subtracted count rate ratio of G2 to G1 (see e.g., Aptekar et al. 2009). We find no spectral evolution in the course of the burst, while the precursor at T$_0$$-$0.460~s is slightly harder. 

There was a total of four high resulution energy spectra collected in the 17 keV$-$12 MeV range with an accumulation time of 64 ms. As there is no evidence of spectral variation, we generated an integrated burst spectrum from T$_0$ to T$_0$+0.256 s to improve statistics. We rebinned the resulting burst spectrum in order to have at least 20 counts per energy bin, and fitted using XSPEC, version 11.3 (Arnaud 1996), in the 17$-$200 keV range since no emission was detected at higher energies. The spectrum is well fitted with an OTTB model: $dN/dE \propto E^{-1} \exp (-E/kT)$ with $kT = 18.0\pm0.9$ ($\chi_\nu^2$ = 0.89 for 17 degrees of fredom [dof]). The total burst fluence is $(6.2\pm0.3)\times10^{-6}$ erg~cm$^{-2}$, and the peak flux is $(3.83\pm0.35)\times10^{-5}$~erg~cm$^{-2}$~s$^{-1}$, as measured at T$_0$ on a 16 ms timescale (both in the 20--200 keV range). Note that errors are at the 90\% confidence level. The corresponding energy release is (1.66$\pm$0.08)$\times$10$^{41}$~erg, and peak luminosity $L_{peak} = 
(1.03\pm0.09)\times10^{42}$~erg~s$^{-1}$, assuming isotropic emission and a distance of 15 kpc (Gaensler et al. 2005). The precursor fluence was $(2.4 \pm 0.3)\times10^{-7}$ erg~cm$^{-2}$, enough to saturate the operating PCA units.

\begin{figure}
\vspace{0.2in}
\begin{center}$
\begin{array}{c}
\includegraphics[scale=.5]{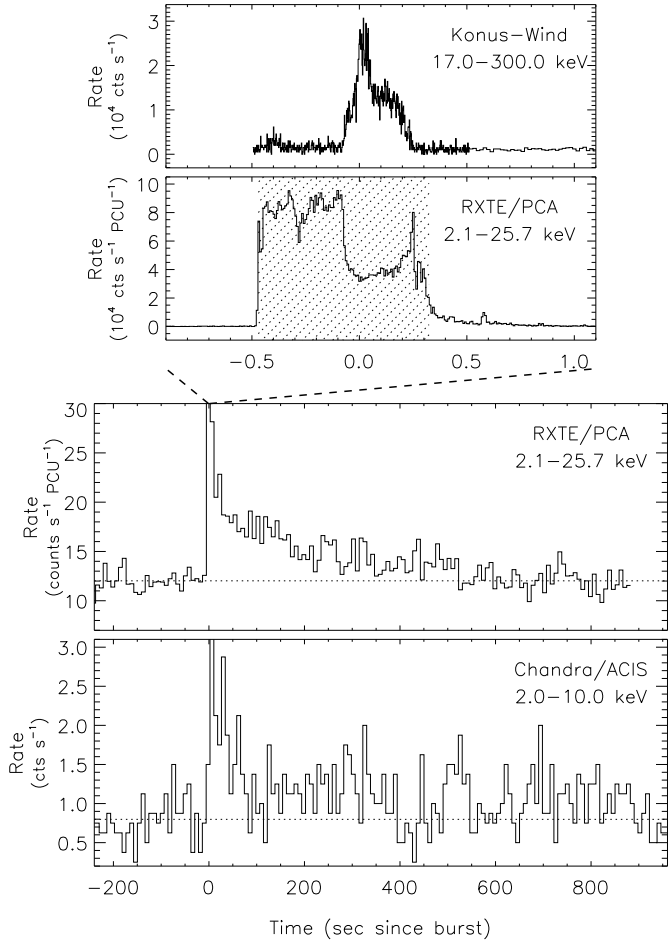}
\end{array}$
\end{center}
\vspace{0.1in}
\caption{Time profile of the event on 22 June 2004, as seen with Konus-Wind
(top panel), and the RXTE/PCA (second panel from top). Here $t=0$ corresponds 
to the Konus-Wind trigger time, i.e., 19:30:09.587 UT. The hashed time span indicates the interval during which the PCA was saturated due to the intensity of the event. The two lower panels show the extended tail of the event in 8 s time 
bins as observed with the RXTE/PCA (second panel from bottom) and with
the Chandra/ACIS (bottom panel). \label{fig:tprof_tail1}}
\end{figure}

({\it ii}) RXTE: We initiated our target of opportunity (ToO) program to 
observe \xsrc with RXTE at the onset of the burst active episode on 
2004 February 15. We detected the first intermediate SGR burst with an extended 
tail among 67 events recorded over a 13.75 ks observation on June 22 (ObsID: 
90073-02-06-01) with the RXTE/Proportional Counter Array (PCA; 2$-$60 keV) 
(see Figure \ref{fig:tprof_tail1}, second panel from top). The brightness 
of the main peak saturated the operating PCA units, resulting in unusable data. 
The total duration of the main event is $\sim$1.3 s, while the tail extends up to about 530 s (Figure \ref{fig:tprof_tail1}; third panel from top). The event was also detected in the simultaneous observations with the RXTE/High Energy X-ray Timing Experiment (HEXTE; 15$-$250 keV). Similar to the PCA data, the extremely high count 
rates of the early part of the event saturated the HEXTE data. In the latter 
data set, the tail extends only up to 1.42 s after the onset, beyond which 
the count rate is consistent with the background.  

({\it iii}) Chandra: We performed simultaneous RXTE and CXO observations of 
\xsrc on 2004 June 22. We used the Advanced CCD Imaging Spectrometer (ACIS) 
in continuous clocking mode for 20.2 ks with the source placed on the S3 
back-illuminated CCD. Inspection of the CXO data revealed detection of the 
same SGR burst recorded with RXTE and KONUS. The main peak of the event 
lasted $\sim$1.5 s in the ACIS energy range of 2$-$10 keV, however, the 
instrument was severely piled-up due to the very high count rate. The average source count rate following the burst was 1.02$\pm$0.14 counts/s -- higher than the count rate prior to the burst (0.81$\pm$0.10 counts/s). The tail was detected until the end of the CXO observation, roughly 900 s after the burst onset (Figure \ref{fig:tprof_tail1}, bottom panel). The source count rate approached the pre-burst level at the end of the CXO observation, but the true tail duration remains uncertain due to the near coincidental termination of the CXO observation.

\subsection{2004 October 17 event} 

\begin{figure}[!h]
\vspace{-0.6in}
\centerline{
\includegraphics[scale=.5]{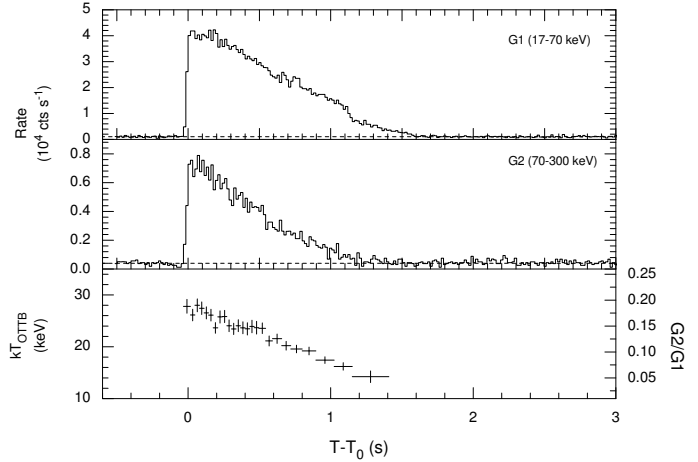}}
\vspace{-0.2in}
\caption{Konus-Wind light curve of the 2004 October 17 intermediate burst in the 17$-$70 keV band (top panel), and in the 70$-$300 keV band (middle panel). Spectral evolution of the burst emission in terms of evaluated temperature of an OTTB model from the hardness ratio (bottom panel). \label{fig:kw_ib2}}
\end{figure}

({\it i}) Konus-Wind: The instrument was triggerred by a bright \xsrc event on
2004 October 17 at 06:36:11.551 UT (Golenetskii et al. 2004b). The T$_{90}$ duration of the burst in the 17-300~keV band was 1.080$\pm$0.032~s (the total duration was about 1.6046~s). We show the Konus$-$Wind light curves in the 17$-$70 and 70$-$300 keV energy bands in Figure~\ref{fig:kw_ib2}. We find a clear hard-to-soft spectral evolution in the course of the burst as evaluated using the hardness ratio (see Figure~\ref{fig:kw_ib2}, bottom panel). 

There were six energy spectra accumulated in the full 17 keV$-$12 MeV Konus-Wind passband during the burst. We combined them into three to improve statistics. The combined spectra are again well described with an OTTB model, with the best fit kT value evolving from 
$25.6_{-1.0}^{+0.8}$ keV in the first 0.256 s interval, to 
21.2$\pm$0.6 keV (0.256$-$0.768 s) and to 
$17.7_{-1.1}^{+1.3}$ keV (0.768$-$1.6 s). 
The total fluence and the 16-ms peak flux in the 20$-$200 keV range are $(5.03\pm0.10)\times10^{-5}$~erg~cm$^{-2}$ and $(6.76\pm0.44)\times10^{-5}$~erg~cm$^{-2}$~s$^{-1}$, respectively. 
We calculate the corresponding isopropic energy released in this event as 
$E = (1.35\pm0.03)\times10^{42}$~erg, and the peak luminosity as
$L_{peak} = (1.81\pm0.12)\times10^{42}$~erg~s$^{-1}$ (at 15 kpc). Events at 11.3 s and 157.2 s after the trigger were also detected, with 20$-$200 keV fluences of $(1.7\pm0.4)\times10^{-7}$~erg~cm$^{-2}$ and  
$(5.6\pm1.7)\times10^{-7}$~erg~cm$^{-2}$, respectively.

\begin{figure}
\vspace{-0.1in}
\begin{center}$
\begin{array}{c}
\includegraphics[scale=.5]{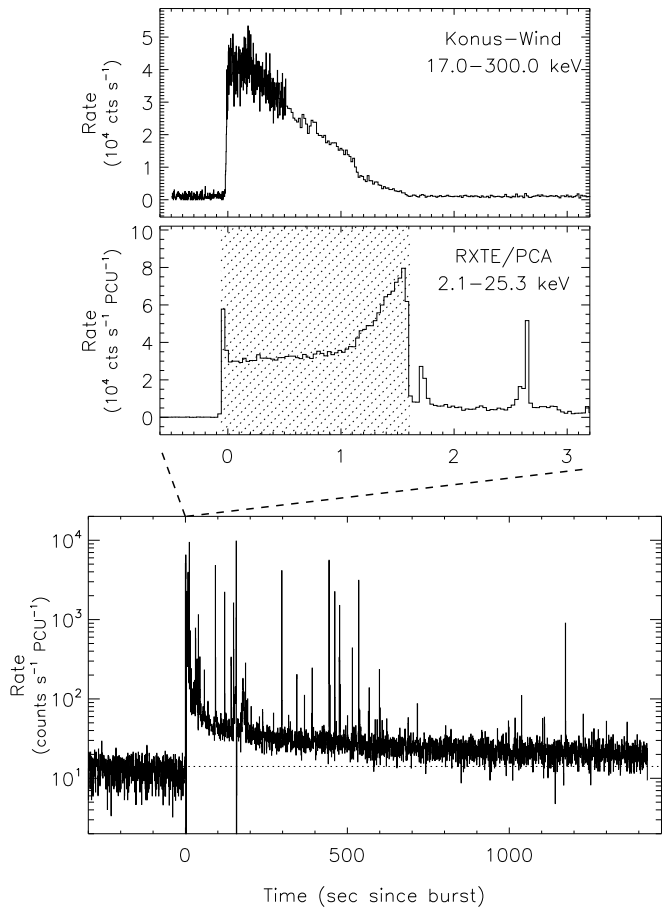}
\end{array}$
\end{center}
\vspace{0.0in}
\caption{Time profile of the event on 2004 October 17 as seen with Konus-Wind
(top panel) and the RXTE/PCA (middle panel). The hashed interval indicates the time span during which the PCA was saturated due to the intensity of the event. The bottom panel shows the extended tail of this event in 1 s time bins. Note that the vertical axis of the bottom panel is logarithmic to accommodate the burst peaks as well as the underlying tail emission. The rate drop-outs at 11.3 and 157.2 s are due to the intense events also detected with Konus-Wind. \label{fig:tprof_tail2}}
\end{figure}

({\it ii}) RXTE: We observed \xsrc with RXTE for 5.7 ks on 2004 October 17 
as a part of our SGR spin monitoring campaign (ObsID: 90074-02-12-00). We found that the triggerred event seen in the Konus-Wind data is also detected with RXTE during the last orbit. The operating PCA units (0,2 and 3) were again saturated for $\sim$1.6 s (see Figure \ref{fig:tprof_tail2}) due to the very high count rates during the main pulse. In the remainder of the RXTE orbit, the count rates stayed above the pre-burst background level (see Figure 2, bottom panel) clearly indicating that the tail emission lasted at least until the end of the orbit. The source was extremely active in this period with a total of 167 short SGR bursts emitted during the tail, as identified with our burst search algorithm. In particular, the relatively bright events at 11.3 s and 157.2 s, that are also seen with Konus-Wind, saturated the detectors as well (see the rate drop out in the bottom panel of Figure \ref{fig:tprof_tail2}). 

In the coexisting HEXTE data we observed 43 short events with the main event and the tail emission, albeit, at much lower significance compared to that of the PCA data. Note that the HEXTE observations were performed in non-rocking mode, and as a result there are no sky background data available for the HEXTE observations.

\section{Data Analysis and Results} 

In the following we present the analysis of the tail emission data for 
both events in RXTE and CXO. Note that the main part of the emission is saturated and piled-up, rendering spectral analysis useless for both bursts and, thus, excluded in our analysis. 

\subsection{Spectral Analysis}

\subsubsection{2004 June 22 Event}
 
As seen in Figure 1 (second panel from bottom), the burst tail  emission
detected with the PCA returns to the pre-burst level at about 600 s after the
burst. We determine the T$_{90}$\footnotemark \footnotetext{The duration over
which 90\% of the tail emission has been recorded.} duration of the tail as
528 s (Figure 3; top panel). We first extracted the time-integrated energy
spectrum during this interval using the PCA Standard2 data  (with 16 s
accumulation time) from all layers of the operating PCUs (\# 0,2,3).

To estimate the PCA background spectrum accurately, we adopted the following 
method. We first extracted a background spectrum using Standard2 data within 
the time interval of 900 -- 50 s before the event. Then we generated an 
instrumental background spectrum using the data created with pcabackest 
(a HEASOFT utility) and subtracted it from the same pre-burst time 
interval. We then modelled the net pre-burst (i.e., the X-ray sky) spectrum with an absorbed power law plus a Gaussian line. The resulting best fit spectral parameters (reduced-$\chi$$^2$ = 0.91) represent the spectral contribution of the SGR, of other point X-ray sources in the 1$^\circ$ (FWHM) PCA field of view and of the diffuse emission from the galactic ridge. While the X-ray sky spectrum would not vary over the course of $\sim$1000 s, the instrumental background spectrum might due to the orbital position of the spacecraft. Therefore, we generated an instrumental background spectrum for the tail interval, again using the data created with pcabackest. The sum of the net pre-burst (X-ray sky) spectrum and the instrumental tail background spectrum, constitutes our combined background spectrum during the tail. 

\begin{figure}[!h]
\vspace{-0.3in}
\begin{center}$
\begin{array}{c}
\includegraphics[scale=.48]{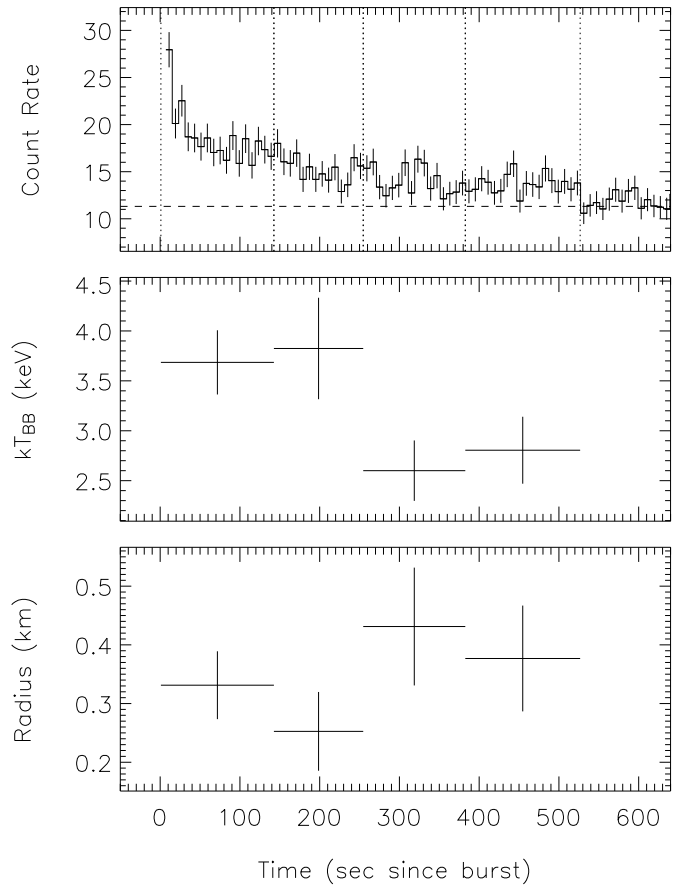}
\end{array}$
\end{center}
\vspace{0.1in}
\caption{{\it top panel:} Time profile of the event on 2004 June 22 in the 
2$-$25 keV range as seen with the RXTE/PCA. The vertical dotted lines indicate
the time intervals over which time resolved spectroscopy was performed.
{\it middle panel:} The variation of the blackbody temperatures and 
{\it bottom panel:} the variation of the corresponding emitting region radii. 
\label{fig:aglow1}
}
\end{figure}

We modeled the PCA tail spectrum (2.5$-$25 keV) with an absorbed blackbody 
(BB), a power law and a bremsstrahlung model (using XSPEC v11 [Arnoud 1996]). 
We fixed the interstellar column density, N$_{\rm H}$ at 7.2$\times$10$^{22}$ 
cm$^{-2}$, that is the value determined 
from the CXO data as reported by \cite{woods07}. We found that the BB
model fits the data best ($\chi$$_{\nu}$$^2$ = 0.73) with kT of 3.4$\pm$0.2 
keV and an estimated emitting radius of 0.30$\pm$0.06 km (for a source distance 
of 15 kpc). The fluence of the entire tail (2.5$-$25 keV) was estimated 
to be (2.9$\pm$0.6)$\times$10$^{-8}$ erg cm$^{2}$. The power law and 
bremsstrahlung models required much higher column densities 
(N$_{\rm H}$ = 12$-$16$\times$10$^{22}$ cm$^{-2}$) to provide acceptable fits, which was in disagreement with the better constrained CXO data results \citep{woods07}. 
We concluded, therefore, that the tail emission was consistent with a thermal 
spectrum. Using the BB spectral fit we estimated the total energy of  $7.9\times10^{38}$ erg, assuming a distance of 15 kpc to the source.

To investigate spectral variations within the tail, we divided the tail 
interval into four segments with almost equal numbers of net tail counts 
(i.e., about 5400 counts per segment and 21780 in total).
We then simultaneously fit all four spectra with an absorbed BB model with fixed N$_{\rm H}$. We found a marginal decrease in the BB temperatures between the first and second halves of the tail  (see also Figure \ref{fig:aglow1}; middle panel). We then repeated the simultaneous fit but  kept the BB temperatures of the first two intervals and the last two intervals linked. This resulted to a temperature of 3.72$\pm$0.27 keV for the first two segments and of 2.68$\pm$0.22 keV for the last two. This temperature difference is significant at the $\sim$3$\sigma$ level, indicating a cooling trend during the tail. The corresponding radii of the emitting region in each of the four subintervals are consistent (within errors) with that of the entire tail spectrum (see Figure \ref{fig:aglow1}; bottom panel).

We also analyzed the Chandra ACIS data of the tail. During the analysis of the ACIS CC mode data, we first identified the pixel corresponding to the centroid of the \xsrc sky position. Using psextract, a CIAO\footnotemark \footnotetext{http://cxc.harvard.edu/ciao} utility, we accumulated the burst tail spectrum from 8 pixels around the centroid for the time interval between 2$-$535 s (determined from the PCA analysis as described above) after the burst onset. The background spectrum was accumulated from the same pixels between 900 to 5 s before the burst. Further, we inspected the light curves of the spectral intervals to 
confirm that there were no short bursts included.

The resulting tail spectrum contained 655 source counts in the 2$-$10 keV 
energy range. Although most of the counts were collected above $\sim$3 keV 
we used the entire spectral range in our fits with BB, power law and 
bremsstrahlung models, all attenuated by interstellar absorption. Since 
our data do not extend below 2 keV, we could not constrain the 
interstellar hydrogen column density (N$_{\rm H}$) and we, therefore, 
fixed N$_{\rm H}$ at 7.2$\times$10$^{22}$ cm$^{-2}$. The BB model again provides the 
best fit with a temperatue of (3.8$\pm$1.3) keV and a emitting region 
radius of (0.36$\pm$0.11) km. The power law model also provides a 
statistically acceptable fit to the data, but the resulting photon 
index is non-constrained, ranging in a wide range between $-$0.66 and 0.60 
(90\% confidence level). The bremsstrahlung model does not fit the spectrum.

\subsubsection{2004 October 17 Event}

The tail emission of this event clearly remained above background level until 
the end of the RXTE pointing, roughly at 1280 s after the burst trigger. 
Therefore, the duration of the tail is at least this long, and most likely 
longer. Before the spectral analysis, we performed a burst search for short 
events in the tail profile (Figure 2; bottom panel) and removed them from the 
tail spectral integration interval. We then constructed the tail background 
spectrum using the same methodology we described in section 3.1.1. 

We fit the entire tail spectrum in the 2.5$-$25 keV range with a BB, a 
power law and a bremsstrahlung model, again keeping the hydrogen column 
density constant at 7.2$\times$10$^{22}$ cm$^{-2}$. 
None of these basic emission models provided an adequate fit to the data 
(with $\chi$$_{\nu}$$^2$ of 11.0, 7.2 and 7.8, respectively). We obtained 
a reasobable  fit to the data with a BB plus a power law model 
($\chi$$_{\nu}$$^2$ = 1.23 for 48 dof). The best fit 
spectral parameters for the model are kT$_{\rm BB}$ = 3.0$\pm$0.2 keV
and photon index, $\gamma$ = 1.2$\pm$0.1.
The total fluence of the tail in the 2.5$-$25 keV band is (4.7$\pm$0.3)
$\times$10$^{-7}$ erg cm$^{2}$, which corresponds to an isotropic total energy of 1.3$\times$10$^{40}$ erg.

\begin{figure}
\vspace{-0.3in}
\begin{center}$
\begin{array}{c}
\includegraphics[scale=.48]{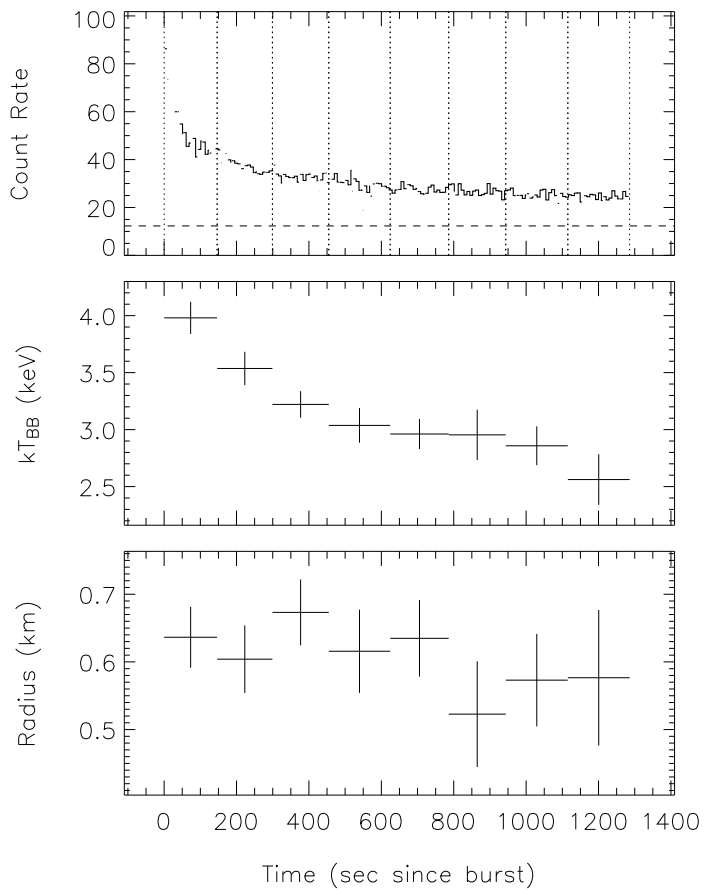}
\end{array}$
\end{center}
\vspace{0.3in}
\caption{({\it top panel:}) Time profile of the event tail on 2004 October 
17 in the 2$-$25 keV range as seen with the RXTE/PCA. The vertical dotted lines 
indicate the time intervals over which time resolved spectroscopy was performed. 
The gaps in the profile correspond to the intervals where short SGR 
bursts were removed from the tail data. ({\it middle panel:}) The variation of the BB temperature and ({\it bottom panel:}) the corresponding emitting region radius. 
\label{fig:tspec_tail2}
}
\end{figure}

We also performed time resolved spectroscopy to search for spectral variations throughout the event tail. We split the tail interval into 8 segments of nearly equal counts of about 16000 counts each. We then simultaneously fit all 8 spectra with a BB plus power law model. We found that the photon indices in all segments were consistent with each other within errors. Therefore, to constrain other spectral parameters better, we subsequently forced all spectra to have a common photon index while we let their normalizations vary. The model adequately fits all data intervals simultaneously ($\chi$$_{\nu}$$^2$ = 0.83 for 399 dof) with a common power law index of $\gamma$ = 1.58$\pm$0.11. We find that the BB temperature clearly decreases with time (see Figure 4; middle panel), from 3.98$\pm$0.14 keV in the first sub-interval to 2.56$\pm$0.22 keV in the last (i.e., a 5.4 $\sigma$ change in about 1000 s). In Figure 5, we show the evolution of the fluxes in both spectral components ($2.5-25$ keV), as well as the ratio of the BB flux to the power law flux during the entire tail interval. We clearly see that the initial part of the burst tail is dominated by the BB flux, while, as the tail progresses, the non-thermal power law component takes over, indicating a clear cooling trend. 

\begin{figure}
\vspace{-0.2in}
\begin{center}$
\begin{array}{c}
\includegraphics[scale=.48]{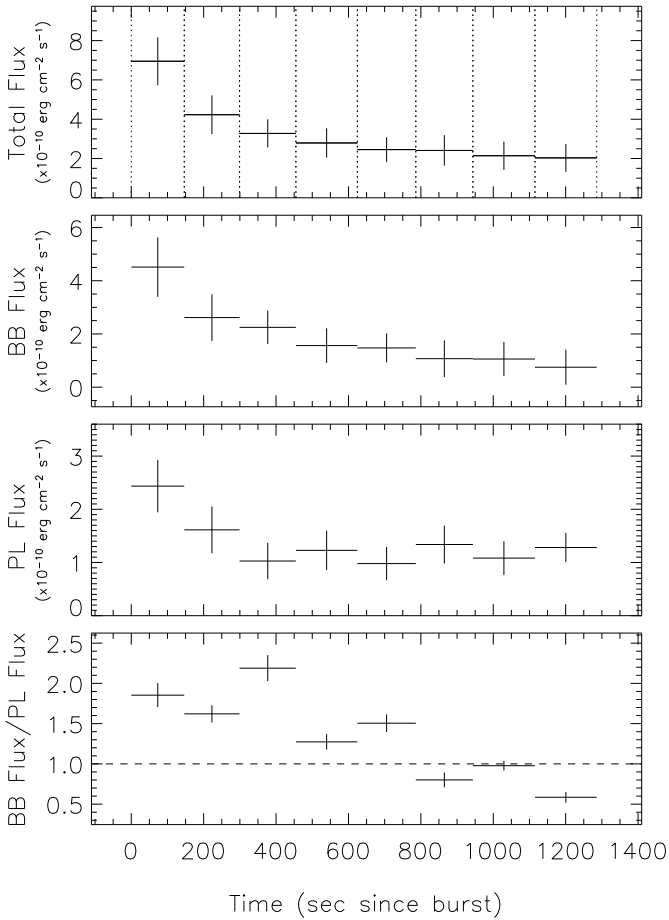}
\end{array}$
\end{center}
\vspace{0.3in}
\caption{The total unabsorbed flux of the event tail on 2004 October 17 
in the 2.5$-$25.5 keV range ({\it top panel}), the unabsorbed flux of the BB
component ({\it second panel from top}), the unabsorbed flux of the power 
law component ({\it third panel from top}), and the ratio of the unabsorbed 
BB flux to that of the power law ({\it bottom panel}). 
\label{fig:tflux_tail2}
}
\end{figure}

We have also analysed the RXTE/HEXTE data to determine the high energy 
behavior of the tail emission. Due to the fact that the HEXTE observation 
was performed in non-rocking mode, we did not have simultaneous sky 
background estimates. We extracted, therefore, a background spectrum from the previous orbit (96 min earlier) for the same time span (Cluster A only). The resulting HEXTE tail spectrum was significant up to $\sim$40 keV and is well represented (in $15-40$ keV) by a single power law with an index of 1.9$\pm$0.2.

\subsection{Temporal Analysis} 

\subsubsection{Burst Induced Changes in Pulsed Intensity}

There were remarkable changes in the persistent X-ray timing and spectral
properties of SGR 1900+14 following the giant flare on 1998 August 27
and the intermediate bursts \citep{woods01,gogus02,lent03}.
To investigate whether these two intense events with extended 
tails from \xsrc had induced similar observable changes, we have estimated the 
pulsed intensity (count rates) before and after each event. \cite{woods07} constructed a comprehensive timing data base of \xsrc, which includes precise pulse frequency ephemerides of the source, surrounding each of the two bursts analyzed here. For the 2004 June 22 event, we extracted time tagged events in the 2$-$10 keV from 1750$-$50 s before and 5$-$535 s after the onset (excluding the times of short burst activity). We then folded the event times with the corresponding spin  ephemeris from Woods et al. (2007) and estimated the RMS pulsed intensity
of the pulse profile before and after the burst as 0.21$\pm$0.05 
counts s$^{-1}$ and 0.46$\pm$0.11 counts s$^{-1}$, respectively. 

Similarly, for 2004 Octover 17 event, we have extracted events from 1750$-$50 s
before and 50$-$1280 s after the event, again excluding the intervals of short
burst activity. We calculated the RMS pulsed intensities as 0.15$\pm$0.08
counts s$^{-1}$ and 0.33$\pm$0.06 counts s$^{-1}$ for before and after this
event. 

\subsubsection{Search for High Frequency QPOs in the Tails}

High frequency quasi periodic oscillations (QPOs) were found in the oscillating
tails of the giant flares from \xsrc and SGR 1900+14 (Israel et al. 2005,
Strohmayer \& Watts 2005, Watts \& Strohmayer 2006). Motivated from these 
findings, we searched the extended tails of the \xsrc bursts for the presence 
of such quasi-periodic behavior by computing power spectra for each 1.89 s 
(0.25 of spin cycles) data segment in the 2$-$10 keV and 10$-$30 keV ranges and 
averaging all available power spectra. We found that average power spectra 
for both tails in the 2$-$10 keV and 10$-$30 keV bands are consistent with 
that of Poisson counting statistics; we estimate the 3$\sigma$ upper limits 
on the RMS amplitudes of QPOs in the 10$-$1024 Hz range for the 2004 June 22
burst tail as 4.4\% and 3.2\% in the 2$-$10 keV and 10$-$30 keV ranges, respectively. The two upper limits for the October 17 burst tail are 4.1\% and 4.0\%, respectively. Therefore, we conclude that there is no evidence of QPOs in these tails. We also found from the dynamic power spectral analysis that there is no evidence for any time or phase dependent timing features in either of these tails.

\subsubsection{Search for Pulse Phase $-$ Burst Time Relations}

\begin{figure*}[t]
\vspace{-0.2in}
\begin{center}$
\begin{array}{cc}
\includegraphics[scale=.43]{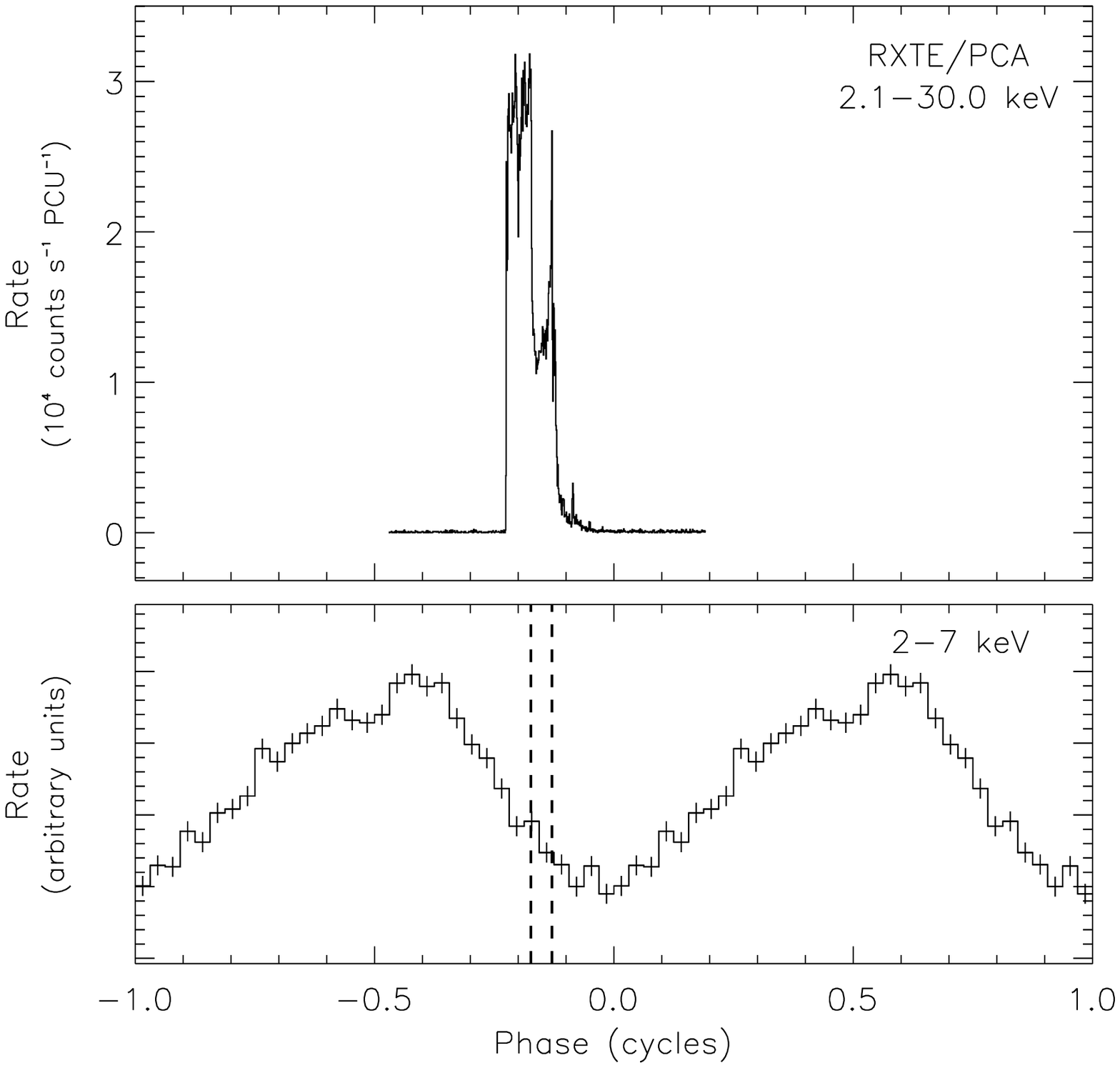} &
\includegraphics[scale=.43]{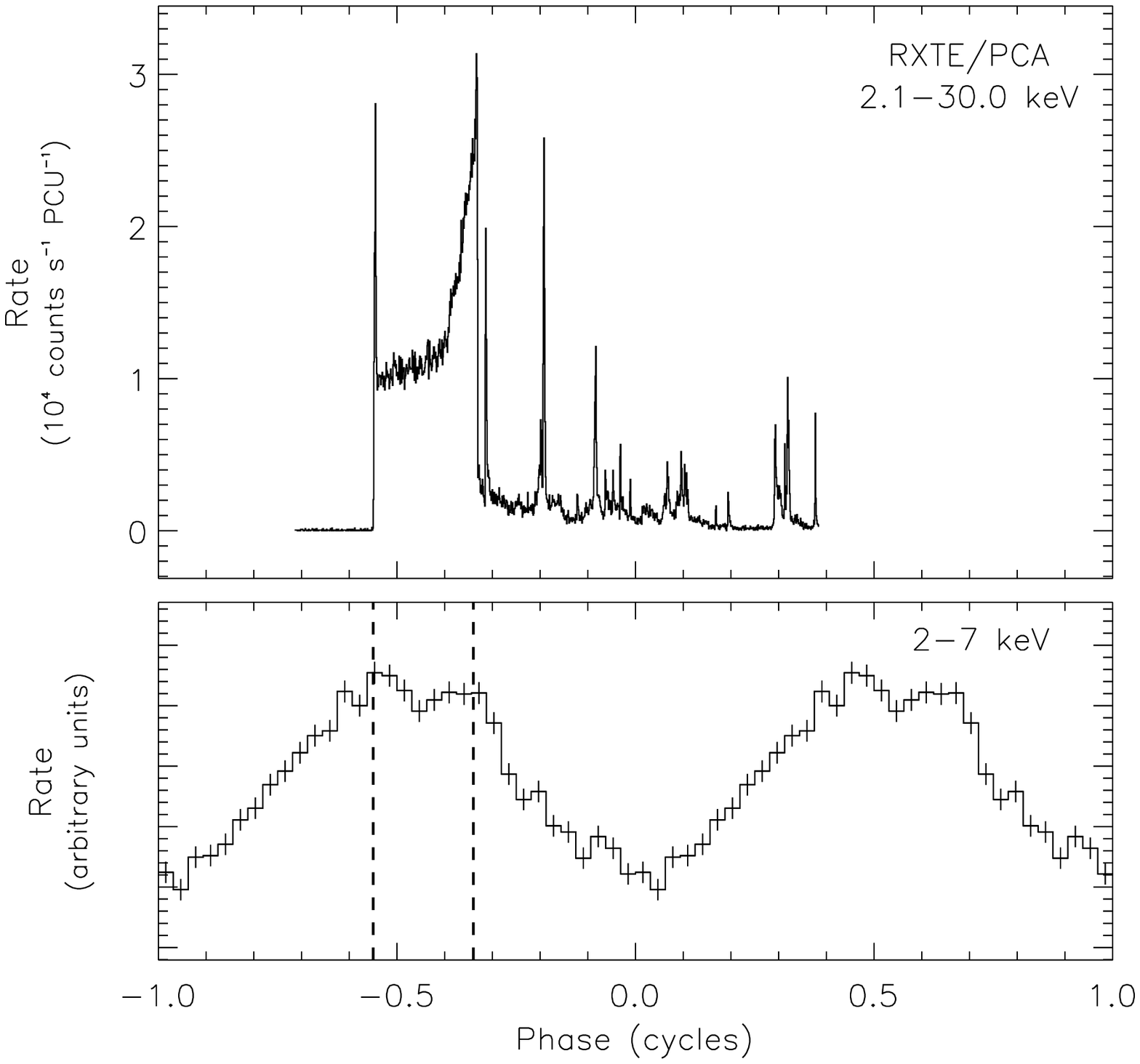}
\end{array}$
\end{center}
\vspace{-0.2in}
\caption{{\it Left Panel:} Phase history of the 2004 June 22 burst as obtained folding the PCA data in the 2.3$-$30 keV band with the phase ephemeris given by Woods et al. (2007) (top) and the average pulse profile of the source in the 2$-$7 keV range events collected during the time span of about two months centered nearly at the time of the burst (bottom). The vertical dashed lines indicate the start and end times of the burst as seen with the Konus-Wind; {\it Right Panel:} Same as the left panel for the 2004 October 17 burst.
\label{fig:bursts_phases}
}
\end{figure*}

To determine over which part of spin phase these bursts took place, we transfered the Konus-Wind times to the solar system barycenter and determined the spin phases of each event using their corresponding pulse frequency ephemerides. We found that 
the main peak of the event on 2004 June 22 (2004 October 17) starts at about 
spin phase, $\phi$=0.75 ($\phi$=0.45) and spans about 4.4\% (21\%) of the 
spin cycle as shown in the lower panels of Figure \ref{fig:bursts_phases}. We also determined the spin phases of the main peaks of the two bursts as seen with the PCA and present them in Figure \ref{fig:bursts_phases}. For comparison, pulse phase behavior of energetic bursts from XTE J1810$-$197 \citep{woods05}, 1E 1048.1$-$5937 \citep{gavr06} and SGR 0501+4516 \citep{watts10} were investigated, but no particular spin phase preference was found.

With the precise pulse frequency ephemeris, we have also checked the spin phases of each of the 167 bursts during the tail of the intermediate burst of 2004 October 17, and found that they are consistent with being distributed uniformly in phase showing no preference for any particular spin phase. 

Further, we looked at the occurence times of the 167 short bursts found in  the
tail of the 2004 October 17, to determine whether they show any intrinsic
periodicity of their own. We employed the Z$_1$$^2$ statistic (Buccheri et al. 
1983) using the burst peak times. The resulting power spectrum consists of a 
single QPO-like feature centered at 7.3 mHz ($\sim$136 s). Closer inspection of the tail revealed that there are two  clusters of events; one at the
very beginning and the other one around  130$-$140 s into the tail, therefore,
the feature in the power spectrum was  likely due to the inter-spacing of
events in these two clusters. We conclude  that there is no periodic behavior
in the occurrence times of the short bursts in the tail.

\section{Discussion} 

We have identified two \xsrc bursts during the source's extremely active episode 
in 2004, that exhibited extended tail emissions similar to the ones reported 
from SGR~$1900+14$ \citep{lent03}. These events happened roughly at 6 and 
1.5 months prior to the SGR Giant Flare on 2004 December 27. Earlier, 
G\"o\u{g}\"u\c{s} et al. (2000) identified 290 short bursts in the RXTE 
observations of \xsrc during its 1996 active episode. Several of these 
events were longer than average (1$-$2 s) but none displayed any extended 
tail emission. 

There are characteristic differences between two bursts we investigated here: the 2004 October 17 bursts is among the events with the highest energy fluence detected during the 2004 active episode of \xsrc and shows clear spectral evolution. The fluence of the 2004 June 22 burst, on the other hand, was not as high and did not exhibit any spectral variation during the course of the event. Due to the fact that both events displayed extended tails following them, the condition for an extended X$-$ray afterglow emission to follow cannot be singly attributed to the energetics of the burst. 

SGR~$1900+14$ emitted two strong bursts on 1998 August 29 and 2001 April 28 
with extended tail emission (durations $\sim$8000 s and $\sim$5000 s, 
respectively). Lenters et al. (2003) studied these event tails extensively; 
they report that the burst fluences (25$-$100 keV) are 1.9$\times$10$^{-5}$ 
erg cm$^{-2}$ and 8.7$\times$10$^{-6}$ erg cm$^{-2}$, respectively. To 
directly compare these to the \xsrc events, we used the Konus-Wind spectral 
results of the main pulse to estimate the fluences of the latter between 
$25-100$ keV; these are 3.6$\times$10$^{-6}$ erg cm$^{-2}$ and 
4.3$\times$10$^{-5}$ erg cm$^{-2}$, for the events of 2004 June 22 and 
October 17, respectively. Therefore, the main burst energetics of these intermediate events from both SGRs are quite similar in the energy range where most of their photons are emitted ($25-100$ keV). 

Lenters et al. (2003) also found that there was a significant correlation 
between the SGR~$1900+14$ event tail fluences in the 2$-$10 keV range and 
the leading main burst fluences in 25$-$100 keV. They estimated the former 
to be about 2\% of the latter fluence.  To check whether a similar 
correlation is present in the \xsrc events, we determined the ratios of their 
tail fluences in the 2$-$10 keV range (1.2$\times$10$^{-8}$ erg cm$^{-2}$ 
and 2.7$\times$10$^{-7}$ erg cm$^{-2}$) to these in their main events. These 
are 0.34\% and 0.63\% for the June 22 and October 17 events, respectively. Although we examined here only two events, we note that these ratios for the \xsrc bursts are both of the same order of magnitude and 3$-$6 times smaller than those of the SGR~$1900+14$ bursts. 

The tail spectra of the two events from SGR~$1900+14$ also included thermal 
components; the dimmer one (2001 April 28 event) was exclusively thermal and 
consistent with a BB temperature of 2.35$\pm$0.05 keV, while the brighter 
one (1998 August 29) had both a thermal (kT = 2.2$\pm$0.2 keV) and a 
non-thermal (power law index of 2.1$\pm$0.2) component (Lenters et al. 2003). 
Similarly, we find that the dimmer \xsrc event tail can be described by a 
single thermal component with a temperature of 3.4$\pm$0.2 keV, while the 
brighter one required two spectral components (kT = 3.5$\pm$0.3 keV and 
photon index = 1.6$\pm$0.1). The BB temperatures obtained from the two 
\xsrc events are consistent with each other, and significantly larger than 
those of the SGR~$1900+14$ events. The dependance of the power law component 
on the burst tail intensity can also be demonstrated from Figure \ref{fig:tflux_tail2}: There we see that the power law component dominates at the later--dimmer stages of the tail, which are only visible in brighter events. As the events become dimmer themselves, only the BB spectral component can be seen in the data. 

\citet{espo07} performed a time resolved spectral analysis 
of the first 7.5 hours of the afterglow emission following the 2001 April 18
flare from SGR 1900+14. They found that an additional BB component was 
required to significantly better fit the afterglow spectrum. They obtain the
best fit parameters by keeping the radius of the BB emitting region constant
at 1.6 km and allowing the BB temperature vary. The temperature 
of the additional BB component gradually declines from 1.23 to 0.92 keV over 
the course of 7.5 hours. They concluded similarly that the cooling thermal 
emission from a constant region could account for the additional spectral 
component.

Recently, \citet{kaneko10} identified a unique enhancement of hard X-ray emission following an intense burst from SGR J1550$-$5418 at the onset of its major outburst episode in 2009. The enhanced emission showed clear pulsations with a period consistent with the spin period of the source. Detailed temporal study revealed that the pulsation was strongest around the peak of the enhancement time-wise and in 50--74\,keV energy-wise with a pulsed fraction of $\sim$55\%. Hard X-ray spectra of the enhancement were well described with a two-component model: a power law with a BB. While the temperature of the BB ($\sim$17\,keV) remained constant throughout the enhancement period, the BB component was most evident around the peak time, where the pulsation was also strongest. The estimated total energy emitted during the enhancement (i.e., the afterglow) was $2.9 \times 10^{40}$\,erg, of which the BB accounted for 19\%. Based on the BB flux, the emitting radius of the thermal component could be as small as 0.12 km. One possible interpretation is that a small region on the surface of the neutron star is likely heated by hot plasma confined by the twisted magnetic field, in which case the energy can be dissipated as the magnetic field untwists \citep{belo09}. 

The pulsed amplitudes of the persistent X-ray emission from SGR 1900+14 during the extended tails were remarkably larger than before the bursts. In fact, the periodic oscillations from SGR 1900+14 were clearly visible in the decaying tail of both intermediate events \citep{ibra01,lent03}, as well as during the enhanced hard X-ray emission of SGR J1550$-$5418 \citep{kaneko10}. In the case of \xsrc events, the changes in the pulsed amplitude in the tails of intermediate bursts were not significant. Moreover, the fraction of energy released in the tail (afterglow) relative to that released during the main the burst (prompt energy release) differs by about an order of magnitude between SGR 1900+14 and \xsrcnos. Although the spectral properties of the \xsrc extended tails are similar to those of the burst afterglows recorded from SGR 1900+14 and from SGR J1550$-$5418, the constancy of the pulsed amplitude in \xsrc and the difference in the fractional energy release indicate intrinsic differences in burst-induced crystal heating (and cooling) mechanisms in these three sources.

We also note here another significant difference between the persistent emission behaviors following Giant Flares from two different sources. The energy released during the 2004 December 27 Giant Flare from \xsrc was at least two orders of magnitude larger than the 1998 August 27 Giant Flare from SGR 1900+14. However, the persistent X-ray flux of \xsrc resumed quickly toward its long term value in the aftermath of the flare \citep{rea05,woods07}, while the Giant Flare of SGR 1900+14 was followed by a long lasting ($\gtrsim$100 days) flux enhancement \citep{woods01}. In the light of these facts, we suggest that \xsrc can efficiently cool by radiation (and baryonic material release \citet{gaensler05}) during the burst (prompt release); most likely also comparatively less energy is imparted to the deep crustal heating than was the case for SGR$1900+14$. An additional possibility for the lack of significant pulsed intensity increase in \xsrc may be the geometry of the emitting areas: the heated crustal surface might be close to the rotation axis, which could be nearly aligned to the magnetic axis of the neutron star. This possibility is indirectly supported by the fact that the main peak of the 2004 October 17 intermediate burst encompasses the peak of its pulse profile.

\acknowledgments 

E.G. and Y.K. acknowledge the support from the Scientific and Technological Council of Turkey (T\"{U}B\.ITAK) through grant 109T755. The Konus-Wind experiment is supported by a Russian Space Agency contract and RFBR grant 09-02-00166a. V.P. acknowledges support from the Ministry of Education and Science of Russian Federation (contract 11.G34.31.0001 with SPbSPU and Leading Scientist G. G. Pavlov).


\begin{thebibliography}{}

\bibitem[Aptekar et al. (2010)]{apte10} Aptekar, R. L. et al. 2009, ApJ, 698, L82

\bibitem[Arnaud (1996)]{arn96} Arnaud, K.A., 1996, ADASS V, eds. Jacoby G. and Barnes J., p17, ASP Conf. Series volume 101

\bibitem[Beloborodov(2009)]{belo09} Beloborodov, A. M. 2009, ApJ, 703, 1044

\bibitem[Buccheri et al.(1983)]{bucc83} Buccheri, R., et al. 1983, A \& A, 128, 245

\bibitem[Eneto et al.(2010)]{eneto10} Eneto, T. et al. (2010), PASJ, 62, 475

\bibitem[Esposito et al.(2007)]{espo07} Esposito, P. et al. (2007), A \& A, 461, 605

\bibitem[Feroci et al.(2003)]{feroci03} Feroci, M. et al. 2003, ApJ, 596, 470

\bibitem[Frederiks et al.(2007)]{fred07} Frederiks, D., et al. 2007, Astron. Letters, 33, 1

\bibitem[Gaensler et al.(2005)]{gaensler05} Gaensler, B. et al. 2005, Nature, 434, 1104

\bibitem[Gavriil, Kaspi \& Woods (2006)]{gavr06} Gavriil, F., Kaspi, V. \& Woods, P.M. 2006, ApJ, 641, 418 

\bibitem[Golenetskii et al.(2004a)]{gol04a} Golenetskii, S. et al. 2004a, GCN 2633

\bibitem[Golenetskii et al.(2004b)]{gol04b} Golenetskii, S. et al. 2004b, GCN 2823 

\bibitem[G\"otz et al.(1999)]{gotz04} Gotz, D. et al. 2004, GCN 2558

\bibitem[G\"o\u{g}\"u\c{s} et al.(1999)]{gogus99} G\"o\u{g}\"u\c{s}, E., et al. 1999, ApJ, 526, L93

\bibitem[G\"o\u{g}\"u\c{s} et al.(2000)]{gogus00} G\"o\u{g}\"u\c{s}, E., et al. 2000, ApJ, 532, L121

\bibitem[G\"o\u{g}\"u\c{s} et al.(2002)]{gogus02} G\"o\u{g}\"u\c{s}, E., et al.  2002, ApJ, 577, 929

\bibitem[Hurley et al.(2005)]{hurl05} Hurley, K., et al. 2005, Nature, 434, 1098

\bibitem[Ibrahim et al.(2001)]{ibra01} Ibrahim, A., et al. 2001, ApJ, 558, 237

\bibitem[Israel et al.(2005)]{isra05} Israel, G., et al. 2005, ApJ, 628, L53

\bibitem[Kaneko et al.(2010)]{kaneko10} Kaneko, Y., et al. 2010, ApJ, 710, 1335



\bibitem[Kouveliotou et~al.(2001)]{kouv01} Kouveliotou, C., et al. 2001, ApJ, 558, L47

\bibitem[Lenters et al.(2003)]{lent03} Lenters, G.T., et al. 2003, ApJ, 587, 761


\bibitem[Mazets et al.(1999)]{mazets99} Mazets, E. P., et al. 1999, ApJ, 519, L151

\bibitem[Mereghetti (2008)]{mereg08}Mereghetti, S. 2008, Astron Astrophys Rev, 15, 225

\bibitem[Mereghetti et al.(2009)]{mereg09} Mereghetti, S., et al. 2009, ApJ, 696, L74

\bibitem[Rea et al.(2005)]{rea05} Rea, N., et al. 2005, ApJ, 627, L133

\bibitem[Palmer et al.(2005)]{palm05} Palmer, D.M., et al. 2005, Nature, 434, 1107

\bibitem[Strohmayer \& Watts (2005)]{sw05} Strohmayer, T.E. \& Watts, A.L. 2005, 632, L111


\bibitem[Watts \& Strohmayer (2006)]{ws06} Watts, A.L. \& Strohmayer, T.E. 2006, ApJ, 637, L117

\bibitem[Watts et al.(2010)]{watts10} Watts, A.L., et al. 2010, ApJ, 719, 190

\bibitem[Woods et al.(2001)]{woods01} Woods, P.M., et al. 2001, ApJ, 552, 748 

\bibitem[Woods et al.(2005)]{woods05} Woods, P.M., et al. 2005, ApJ, 629, 985 

\bibitem[Woods \& Thompson (2006)]{woods06} Woods, P.M., \& Thompson, C. 2006, in Compact Stellar X-ray Sources, eds. W.H.G. Lewin \& M. van der Klis, Cambridge Astrophysics Series, 39, p.547

\bibitem[Woods et al.(2007)]{woods07} Woods, P.M., et al. 2007, ApJ, 654, 470 

\end{thebibliography}
\end{document}